# High-Redshift Radio Galaxies as a Cosmological Tool: Exploration of a Key Assumption and Comparison with Supernova Results

Ruth A. Daly[*], Matthew P. Mory[*] and Erick J. Guerra[†]

[*]Department of Physics, Penn State University, Berks-Lehigh Valley College, P.O. Box 7009, Reading, PA 19610-6009, USA
[†]Department of Chemistry & Physics, Rowan University, Classboro, NJ 08028-1701, USA

**Abstract.** There are many different approaches to using observations to constrain or determine the global cosmological parameters that describe our universe. Methods that rely upon a determination of the coordinate distance to high-redshift sources are particularly useful because they do not involve assumptions about the clustering properties of matter, or the evolution of this clustering.

Two of the methods currently being used to determine the coordinate distance to high-redshift sources are the radio galaxy method and the supernova method. These methods are similar in their dependence on the coordinate distance. Here, the radio galaxy method is briefly described and results are presented. One of the underlying assumptions of the method is explored. In addition, the method is compared and contrasted to the supernova method. The constraints imposed on global cosmological parameters by radio galaxies are consistent with those imposed by supernovae.

For a universe that is spatially flat with mean mass density $\Omega_m$ in non-relativistic matter and mean mass density 1- $\Omega_m$ in quintessence, radio galaxies alone indicate at 84 % confidence that the expansion of the universe is accelerating at the current epoch. And, independent of whether or not the universe is spatially flat, radio galaxies alone indicate at 95 % confidence that $\Omega_m$ must be less than 0.6 at the current epoch.

## 1. INTRODUCTION

Global cosmological parameters describe the current state of the universe and indicate the future of the universe and are thus of obvious importance. One way to parameterize current values of global cosmological parameters is through the equation of state and mean mass density of the component at the present epoch. Observations of stars, galaxies, active galaxies, and other sources are then used in an attempt to constrain these parameters.

The different methods used to constrain global cosmological parameters have different assumptions and constrain the parameters in different ways. Determinations of the coordinate distance $(a_o r)$ [1] as a means to constrain global cosmological parameters is particularly useful because it allows a direct determination of global cosmological parameters. The method is independent of all aspects of density fluctuations, including the source and growth of these fluctuations. Thus, the method is particularly clean.

The properties of radio galaxies can be used to deterime the coordinate distance to sources with redshifts from zero to two. The method is outlined in section 2, and one of the primary assumptions is explored in detail in secion 2.1. General equations that apply to global cosmological parameters including quintessence in a spatially flat are presented in section 2.2. Supernova and radio galaxy constraints on global cosmological parameters are compared in section 3. Conclusions follow in section 4.

## 2. THE RADIO GALAXY METHOD AND RESULTS

The use of powerful extended FRII radio sources as a cosmological tool is presented and discussed by [2], [3], [4], [5] and [6]. The properties of the sources used for cosmology are explored in more detail by [7].

The use of very powerful FRII radio sources for cosmology is based on three assumptions. (1) Very powerful classical double radio sources are supersonic propagators, and thus the equations of strong shock physics apply. This seems justified based on the radio bridge properties of the sources, [8], [9], and [7]. (2) The average size a given source will have over its lifetime will be close to the average size of the full population of similar sources at the same redshift. This seems justified by the fact that the dispersion in source size at a given redshift is rather small. (3) The total time the AGN will produce highly collimated outflows to drive the growth of the source, $t_*$, is related to the beam power of the source, $L_j$, through the relation $t_* \propto L_j^{-\beta/3}$, where $\beta$ is a parameter whose value needs to be determined.

The third assumption fits in rather nicely with currently popular models of jet production due to the electromagnetic energy extraction from a rotating black hole, and is discussed in some detail below.

### 2.1. Electromagnetic Energy Extraction from Rotating Holes

An AGN that produces two oppositely directed collimated outflows each with beam power $L_j$ for a total time $t_*$ will release a total energy $E_* = 2L_j t_*$. Thus, the relation $t_* \propto L_j^{-\beta/3}$ is equivalent to the relation $E_* \propto L_j^{1-\beta/3}$; this is equivalent to the third assumption described above. The quantities $E_*$ and $L_j$ for a model in which the beam power derives from the electromagnetic extraction of spin energy from a rotating black hole are described by [10]. Blandford [10] shows that the beam power and total energy available are

$$L_j = L_{EM} \sim 10^{45} (a/m)^2 \, B_4^2 \, M_8^2 \text{ erg s}^{-1} \propto (a/m)^2 \, B^2 \, M^2 \tag{1}$$

and

$$E_* = E \sim 5 \times 10^{61} (a/m)^2 \, M_8 \text{ erg} \propto (a/m)^2 \, M \;, \tag{2}$$

for $(a/m) << 1$, where $M$ is the mass of the black hole, $M_8$ is the mass in units of $10^8 M_\odot$, $a$ is the spin angular momentum $S$ per unit mass $M$: $a = S/(Mc)$, c is the speed of light, $m$ is the gravitational radius $m = GM/c^2$, $B$ is the magnetic field strength, and $B_4$ is the magnetic field strength in units of $10^4$ G [10].

An exploration of the properties of very powerful double radio galaxies indicates that $E_* \propto L_j^{1-\beta/3}$ with $\beta = 1.75 \pm 0.25$ as discussed by [4], [5], [11]and [6]. This is consistent with equations (2) and (3) when the magnetic field strength satisfies

$$B \propto M^{(2\beta-3)/2(3-\beta)} \, (a/m)^{\beta/(3-\beta)} \, . \tag{3}$$

For the case $\beta = 1.5$ simplifies beautifully to $B \propto (a/m)$. For the cases $\beta = 1.75$ and 2, the empirical determined relation is consistent with equations (3) and (4) when $B \propto (a/m)^{1.4} \, M^{0.2}$ and $B \propto (a/m)^2 \, M^{1/2}$ respectively.

A comparison of the results published by [5] indicates that a when $L_j \sim 10^{45}$ erg s$^{-1}$, the total energy processed through large-scale jets is $\sim 5 \times 10^5 M_\odot c^2$, so the jets are active for a total lifetime of $\sim 10^7$ yr. Equations 1 and 2 show that the total source lifetime is about $t_* \sim 10^9/(M_8 B_4^2)$, indicating a total lifetime of about $10^7$ yr for $M_8 B_4^2 \sim 10^2$. This is satisfied for $M_* \sim 10$, and $B_4 \sim 3$. In this case, the beam power is $\sim 10^{45}$ erg s$^{-1}$ for $(a/m) \sim (1/30)$, and the total energy is $\sim 5 \times 10^5 M_\odot$. These values for $M_8$, $B_4$, and $(a/m)$ seem quite reasonable. The scaling between variables required by the empirically determine relation between total energy and beam power are given above.

## 2.2. General Cosmological Equations for a Universe with Quintessence

Recent measurements of the cosmic microwave background anisotropy suggest that the universe has zero space curvature (de Bernardis et al. 2000, Balbi et al. 2000). Here, a universe with zero space curvature ($k = 0$)and quintessence is considered. These equations will be applied below to consider the constraints placed on quintessence by radio galaxies. These equations are well known, and can be found in [1], [12], and [14], to name a few. Quintessence is introduced and discussed by [13],[14], and [15].

The deceleration parameter $q_o$ is defined to be $q_o = -(\ddot{a}a)/\dot{a}^2$ evaluated at the present epoch z=0, where $a$ is the cosmic scale factor. This can be re-written $q_o = -\ddot{a}_o/(a_o H_o^2)$; quantities evaluated at z=0 have a subscript 'o', and $H_o = (\dot{a}_o/a_o)$. It is shown in [12] that

$$(\ddot{a}/a) = -\frac{4\pi G}{3} \sum (\rho_i + 3p_i) = -\frac{4\pi G}{3} \sum \rho_i (1 + 3w_i) \tag{4}$$

where $p_i$ is the pressure, $\rho_i$ is the mean mass-energy density, $w_i$ is the equation of state of $i$th component, $w_i = p_i/\rho_i$, with the quintessence term is included in the summation.

Mass-energy conservation of each component implies that

$$\dot{\rho}_i = -3(\rho_i + p_i)(\dot{a}/a) \tag{5}$$

[12]. By definition, the equation of state is $w_i = p_i/\rho_i$, so the eq. 5 implies $(\dot{\rho}_i/\rho_i) = -3(1 + w_i)(\dot{a}/a)$ . When the equation of state $w_i$ does not change with time, the solution to this equation is $\rho_i = \rho_{i,o}(1+z)^{3(1+w_i)}$, where $(1+z) = a_o/a$. Thus, a component with equation of state $w_i$ and present mean mass-energy density $\rho_o$ will have a mean mass-energy density at redshift $z$ of $\rho = \rho_o(1+z)^{n_i}$, where $n_i = 3(1 + w_i)$.

When $k = 0$,

$$(\dot{a}/a)^2 = \frac{8\pi G}{3} \sum \rho_i , \qquad (6)$$

where $\rho_i = \rho_{i,o}(1+z)^n$. It follows that $H_o^2 = (\dot{a}_o/a_o)^2 = (\frac{8\pi G}{3}) \sum \rho_{i,o}$. For $k = 0$, $\sum \rho_{i,o} = \rho_{c,o}$ where $\rho_{c,o}$ is the critical density at redshift zero. So,

$$H_o^2 = \frac{8\pi G}{3} \rho_{c,o} \qquad (7)$$

for $k = 0$. The deceleration parameter $q_o = -\ddot{a}_o/(a_o H_o^2)$ then becomes

$$q_o = (1/2)\sum \Omega_i (1 + 3w_i) , \qquad (8)$$

where $\Omega_i = \rho_{i,o}/\rho_{c,o}$ and equations 4 and 7 have been used.

The universe is accelerating when $q_o < 0$. This can only occur if $(1 + 3w_i) < 0$, or $w_i < -1/3$, which is a necessary but not sufficient condition to have an accelerating universe at the present epoch. When there are only two types of mass-energy controlling the expansion rate of the universe at the the current epoch, quintessence and non-relativistic matter, then the deceleration parameter is $q_o = \Omega_m/2 + \Omega_Q(1 + 3w)/2$. The universe will be accelerating in its expansion when

$$1 + 3w(1 - \Omega_m) < 0 , \qquad (9)$$

which follows since $\Omega_Q = 1 - \Omega_m$; similar equations are presented by [14]. Thus, there are two regions on the $\Omega_m - w$ plane; points in one region represent solutions for which the universe is currently accelerating, while those in the other region represent solutions for which the universe is currently decelerating (e.g. see Figure 1). The curve separating these regions is indicated on Figure 1, which illustrates the constraints obtained using radio galaxies [6]. Note that radio galaxies alone place interesting constraints on $\Omega_m$ and $w$, and these results are consistent with those obtained using other methods (e.g Wang et al. 2000).

The coordinate distance to a source at redshift $z$ follows from the equation

$$\int dr/\sqrt{1 - kr^2} = \int dt/a(t) = (1/a_o) \int (\dot{a}/a)^{-1} dz \qquad (10)$$

[1]. For a spatially flat universe, the left hand side of the equation reduces to the coordinate distance $r$. Equations 6 and 7 imply that $(\dot{a}/a)^2 = (H_o^2/\rho_{c,o}) \sum \rho_{i,o}(1+z)^{n_i}$ or

$$(\dot{a}/a)^2 = H_o^2 \sum \Omega_i (1+z)^{n_i} . \qquad (11)$$

Following [12], $(\dot{a}/a) = H_o E(z)$, where

$$E(z) = \sqrt{\sum \Omega_i (1+z)^{n_i}} . \qquad (12)$$

The coordinate distance to a source at redshift $z$ in a spatially flat universe is (see equation 10)

$$(a_o r) = H_o^{-1} \int_0^z dz/E(z) . \qquad (13)$$

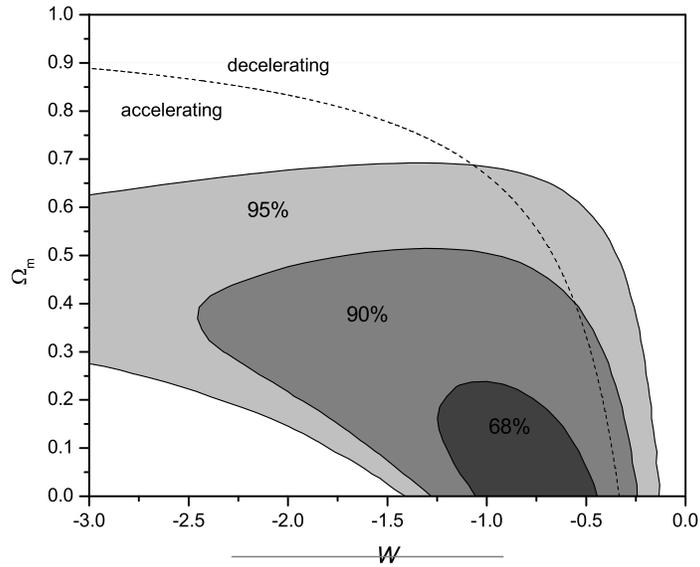

**FIGURE 1.** One-dimensional confidence contours obtain using radio galaxies, assuming a spatially flat universe. The dotted curve separates regions for which solutions represent an accelerating or decelerating universe. Conference participants were interested in values of *w* less than -1.0, so the region shown has been extended to w=-3.0. A similar figure is published by [6].

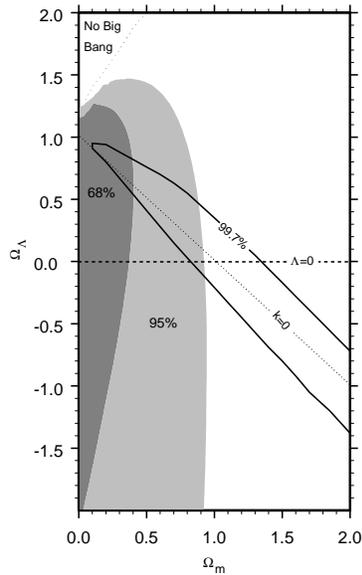

**FIGURE 2.** Constraints from the cosmic microwave background from [16] are indicated by the solid lines, while the two-dimensional constraints from radio galaxies are indicated by the shaded regions. Clearly, these two sets of observations imply a flat, low-density universe.

**TABLE 1.** Comparison Between Supernova and Radio Galaxy Methods.

| Supenovae | Radio Galaxies |
|---|---|
| Type SNIa | Type FRIIb |
| $\propto (a_o r)^{2.0}$ | $\propto (a_o r)^{1.6}$ |
| $0 < z < 1$ | $0 < z < 2$ |
| $\sim 100$ sources | 20 sources (70 in parent pop.) |
| modified standard candle | modified standard yardstick |
| light curve $\Longrightarrow$ peak luminosity | radio bridge $\Longrightarrow$ average length |
| empirical relation | physical relation |
| written in terms of observables | written in terms of physical variables |
| normalized at $z = 0$ | not normalized at $z = 0$ |
| dependent on local distance scale | independent of local distance scale |
| universe is accelerating | $\Omega_m$ is low |
| | universe is acclelerating if k=0 |
| some theoretical understanding | good theoretical understanding |
| well tested empirically | needs more empirical testing |

A spatially flat universe has $\sum \Omega_i = 1$. The components that must be included in equation 12 are those that contribute from redshift zero out to the redshift to which the coordinate distance is being determined. Here, two components are considered: non-relativistic matter with normalized mean mass-energy density at $z = 0$ of $\Omega_m$, and quintessence with normalized mean mass-energy density at $z = 0$ of $\Omega_Q$. Thus, $\Omega_m + \Omega_Q = 1$, and equation 12 becomes $E(z) = \sqrt{\Omega_m (1+z)^3 + (1-\Omega_m)(1+z)^n}$. The coordinate distance is obtained by substituting this into eq. 13 and solving for $(a_o r)$.

## 3. A COMPARISON OF SUPERNOVA AND RADIO GALAXY RESULTS

The supernova and radio galaxy methods are compared in detail by [6]. These results are summarized in Table 1 and Figures 3 and 4.

## 4. CONCLUSIONS

Only a few methods are available at present to constrain global cosmological parameters through the determination of the coordinate distance to high-redshift sources. In this paper the results presented by [6], and [17] are presented and summarized.

## ACKNOWLEDGMENTS

It is a pleasure to thank the organizers of the Coral Gables conference for such an interesting and interactive meeting. In particular I would like to thank Behram Kursunoglu, Sydney Meshkov, Arnold Perlmutter, and Ina Sarcevic. I would like to acknowledge numerous interesting and helpful discussions with conference participants

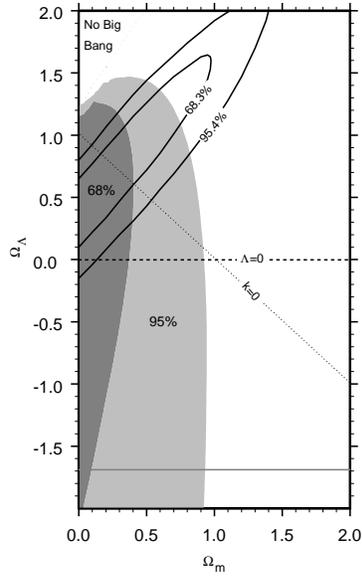

**FIGURE 3.** Constraints from the high-redshift supernova team [20] are indicated by the solid lines, while the radio galaxy results are indicated by the shaded regions; both are two-dimensional constraints.

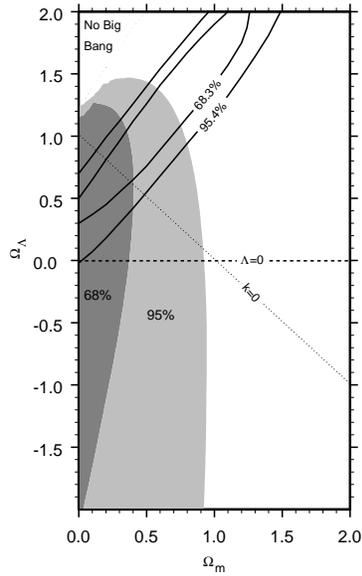

**FIGURE 4.** Constraints from the supernovae cosmology team [19] are indicated by the solid lines, while the radio galaxy results are indicated by the shaded regions; both are two-dimensional constraints.

Philip Mannheim, Paul Frampton, Ina Sarcevic, and Lynn Cominsky, which were particularly valuable. I would also like to thank Chris O'Dea for numerous helpful discussion related to radio sources. This work was supported in part by a National Young Investigator Award Grant number AST-0096077 from the US National Science Foundation, the Berks-Lehigh Valley College of Penn State University. Research at Rowan University was supported in part by teh College of Liberal Arts and Sciences and National Science Foundation grant AST-9905652.